# Carbon nanotube composites for thermal management

M.J. Biercuk[a)], M.C. Llaguno, M. Radosavljevic[b)], J.K. Hyun[c)], and A.T. Johnson[d)]

*Department of Physics & Astronomy and Laboratory for Research on the Structure of Matter, University of Pennsylvania, Philadelphia, Pennsylvania   19104*

J.E. Fischer

*Department of Materials Science & Engineering and Laboratory for Research on the Structure of Matter, University of Pennsylvania, Philadelphia, Pennsylvania   19104*

Single-wall carbon nanotubes (SWNTs) were used to augment the thermal transport properties of industrial epoxy. Samples loaded with 1 wt% unpurified SWNT material show a 70% increase in thermal conductivity at 40K, rising to 125% at room temperature; the enhancement due to 1 wt% loading of vapor grown carbon fibers is three times smaller. Electrical conductivity data show a percolation threshold between 0.1 and 0.2 wt% SWNT loading. The Vickers hardness rises monotonically with SWNT loading up to a factor of 3.5 at 2 wt%. These results suggest that the thermal and mechanical properties of SWNT-epoxy composites are improved, without the need to chemically functionalize the nanotubes.

a) present address: Department of Physics, Harvard University, Cambridge, Massachusetts   02138.
b) present address: IBM T.J. Watson Research Center, Yorktown Heights, New York   10598
c) present address: Department of Applied Physics, Columbia University, New York, New York   10027
d) electronic mail: cjohnson@physics.upenn.edu

Rapid advance in the large-scale synthesis of single wall carbon nanotubes (SWNTs), coupled with their remarkable mechanical properties, are positive signs for application in a host of composite materials. First, the high Young's modulus and strength-to-weight ratio of SWNTs are extremely promising for enhanced strength in mechanical composites. Second, and only more recently appreciated, because phonons dominate thermal transport at all temperatures in carbon materials [1], and since the phonon thermal conductivity ($\kappa$) can be estimated by $\kappa = C_P v_s l$ (where $C_P$ is the specific heat, $v_s$ the speed of sound, and l the mean free path), SWNTs should be ideal for high-performance thermal management [2]. Indeed, theory predicts an extremely high value (6000 W/m-K) for the room temperature thermal conductivity of an isolated SWNT [3]. Measurements of $\kappa(T)$ for bulk SWNT samples agree with predictions for one-dimensional phonon systems [4, 5], while a recent measurement of 3000 W/m-K for the room temperature thermal conductivity of an *individual* multiwall nanotube (MWNT) [6] shows directly that the superior thermal properties of the graphene plane are realized in its related nanomaterials.

Despite this great promise, progress in mechanical composites has proved difficult to date. Although enhanced strength was observed in SWNT-polymethylmethacrylate (PMMA) composites [7], SWNT-epoxy and MWNT-epoxy composites are typically weaker or only slightly stronger than the pristine epoxy [8, 9, 10]. This has been attributed to a weak nanotube-matrix interaction but may also reflect processing difficulties that result in poor SWNT dispersion [11]. We report here our results on the thermal and mechanical properties of SWNT-epoxy composites.

We fabricated nanotube-epoxy thermal composites and measured a thermal conductivity enhancement of more than 125% at 1.0 wt% nanotube loading. Composites loaded with vapor grown carbon fibers (VGCF) at the same mass fraction show considerably weaker enhancement (45%). As is typical for composite materials, both of these measured enhancements are significantly lower than the "law of mixtures" prediction, even taking a conservative estimate of 1000 W/m-K for the SWNT thermal conductivity, and the published value of 1900 W/m-K for the VGCFs. Measurements of the electrical conductivity show a sharp increase by a factor of nearly $10^5$ between 0.1 and 0.2 wt% SWNT loading but not until between 1 and 2 wt% for VGCF composites. This likely indicates easier formation of a percolation network of SWNTs because of their larger aspect ratio. We used micro-indentation testing (Vickers hardness or VH) to evaluate the mechanical properties because of limited material quantity. There is a monotonic increase in VH with SWNT loading that reaches a factor of 3.5 at 2 wt%.

Samples were based on the Shell Chemicals Epon 862 epoxy resin and Air Products Ancamine 2435 dimethane-amine curing agent. Composite samples were loaded with either raw SWNT soot grown by the HiPCO method [12] or vapor grown carbon fibers (VGCF) from Applied Sciences, Inc. The SWNT material contained approximately 15 – 25 wt% Fe catalyst in the form of isolated nanoparticles. Quoted loading values are based on the mass of as-grown SWNT material added to epoxy and have not been reduced to account for the Fe impurities. Transmission electron microscopy revealed that the SWNT material had a broad distribution in tube diameter peaked at 1.1 nm and tube lengths from hundreds of nanometers to several microns. Nanotube bundles were small (3 – 30 nm), and neutron diffraction indicated poor bundle crystallinity [13]. "Pyrograf-III" vapor grown carbon fibers (VGCF) have an average diameter of 200 nm and lengths from several microns to over 10 mm. During sample production (mixing, sonication; see below) these fibers tend to break to lengths below 100 μm. The thermal conductivity of the fibers is 1900 W/m–K at room temperature [14], close to that predicted for SWNTs. Control samples of pristine epoxy were also measured.

Care was taken to disperse carbon material uniformly through the composite. Carbon materials were dispersed ultrasonically for as long as 48 hours in an organic solvent (dichloroethane or N-N dimethylformamide) to promote the formation of a stable suspension. The epoxy resin was subsequently dissolved in the carbon/solvent mixture; high quality dispersal was indicated by the formation of a smooth emulsion. This solution was placed under vacuum to remove trapped air. After degassing, samples were placed on a hotplate at 130 C for an hour to completely evaporate the solvent [15]. The curing agent was added, and the samples cured at room temperature for two to four days followed by a 2 h post-cure bake at 120 C. Loading was studied up to 5 wt%; Fig. 1 shows that the SWNTs were well dispersed in the material on the micrometer scale, with random tube/rope orientation. Samples were black, however, indicating that the dispersal was still not ideal.

From the bulk product we cut samples (typically 1 mm x 1 mm x 2 mm), taking care to exclude regions with macroscopic air inclusions. We verified that different samples cut from the same epoxy composite gave identical results for all measurements. Thermal conductivity was measured from 20 – 300K with a comparative technique [16]. Briefly, a sample is mounted between two constantan rods of known thermal conductivity. A heat current is passed through this thermal "circuit" and then to thermal ground (the sample stage). Temperature drops are measured across each rod and the sample using differential thermocouples, yielding the relative thermal conductivity of the sample. The second standard is used to monitor heat current loss due to radiation and other sources, with care taken to minimize these negative effects. In calibration runs, this method accurately reproduced the known thermal conductivity of numerous different samples of metals and carbon fibers.

For the room temperature data as a function of carbon loading, results from 5 – 10 measurements were averaged for each sample. The thermal conductivity enhancement in SWNT-epoxy samples rises much more rapidly than in samples loaded with VGCF. We observe a 125 % enhancement in κ at 1 wt% SWNT loading while equal loading of VGCF produces but a 45% increase. Data for κ(T) were collected in a closed-cycle helium-cooled system; typical raw data and the enhancement in κ(T) are presented in Fig. 3. From this data we see that SWNT-epoxy composites show a markedly higher κ enhancement at all temperatures than VGCF-loaded samples.

Room temperature electrical conductivity of the composites was measured to characterize the extent of the carbon material network. For both carbon materials, low-loading samples display only a factor of 2 or 3 change in conductivity followed by a sudden jump by more than $10^4$, consistent with the formation of a percolating network through the sample. The percolation threshold, where the sharp onset in conductivity is observed, is between 0.1 and 0.2 wt% for SWNT-epoxy samples, while VGCF requires loading between 1 and 2 wt%. These values are reasonable in light of experimental,[17] numreical,[18] and theoretical[19] work indicating that the percolation threshold in dilute, random rod systems is approximately equal to the inverse of the aspect ration, roughly 1000 for SWNTs and 100 for VGCF. Our measured critical loading for SWNT is four times smaller than that found by others for SWNT-PMMA composites [20]. This difference reflects the fact that their samples were formed by spin coating which preferentially aligns nanotubes in the plane of the sample, normal to the current direction in their experiment. An even more extreme case is presented by SWNT-PMMA fibers of Ref. 7. They were made with a very large draw ratio, leading to high SWNT alignment , [7] and are electrical insulators even at loading as high as 5 wt %. [21] In sharp contrast to these cases, the SWNTs in our samples are oriented randomly, and the percolation network forms at lower loading.

Resistance to plastic flow for HiPCO-epoxy samples was measured using a Tukon Microhardness Tester with a Vickers indenter. Epoxy samples were mechanically polished to ensure a smooth sample surface, and VH measured using a 100 gram load. The VH increased almost linearly from 0.4 for the pristine epoxy to 1.4 at 2 wt% loading.

These experiments demonstrate convincingly that a small fraction of SWNT material can dramatically enhance the thermal properties of an epoxy matrix, and that they are more effective than larger diameter carbon fibers for this purpose. It is likely that SWNTs are superior to VGCFs because their nanoscale diameter and larger aspect ratio enable a more extensive network to form at the same weight loading, as indicated by the lower critical loading for percolation seen in the electrical conductivity measurements. The observed increase in VH is intriguing. First, it implies that SWNT-based thermal composites do not suffer any degradation in mechanical strength, a fact that augurs well for applications. Second, the increase in VH suggests that high-quality SWNT dispersion may be an important factor in enhancing the strength of SWNT-epoxy composites.

Ongoing TEM imaging to characterize the SWNT dispersion on the nanometer scale, and mechanical tests that diretly measure the elastic moduli of the composites will further illuminate this point.

In summary, we demonstrated that SWNT-epoxy composites have significantly enhanced thermal conductivity, and SWNTs are much more effective for this than larger-diameter carbon fibers. We observed an increase in Vickers hardness with increasing SWNT loading that will be the subject of additional investigation. Future improvements might result from chemical functionalization of the SWNTs to enhance the SWNT-matrix interaction and the creation of anisotropic thermal management composites via nanotube alignment.

This work was supported by the US Department of Energy, DEFG02-98ER45701 (MCL and JEF), and the Laboratory for Research on the Structure of Matter (LRSM), an NSF-supported MRSEC, DMR00-79909 (ATJ, MR). MJB and JKH acknowledge the support of the LRSM Research Experience for Undergraduates program.

FIGURE CAPTIONS

Figure 1: Scanning electron micrograph of 1 wt% SWNT-epoxy composite. Randomly oriented nanotube bundles are clearly visible throughout the matrix.

Figure 2: Enhancement in thermal conductivity relative to pristine epoxy as a function of SWNT and VGCF loading.

Figure 3: (a) Thermal conductivity vs. temperature for the pristine epoxy and epoxy with 1 wt% SWNT loading. (b) Enhancement in the thermal conductivity vs. temperature for composites loaded with 1 wt% SWNT and VGCF. The SWNT-epoxy has a larger enhancement at all temperatures.

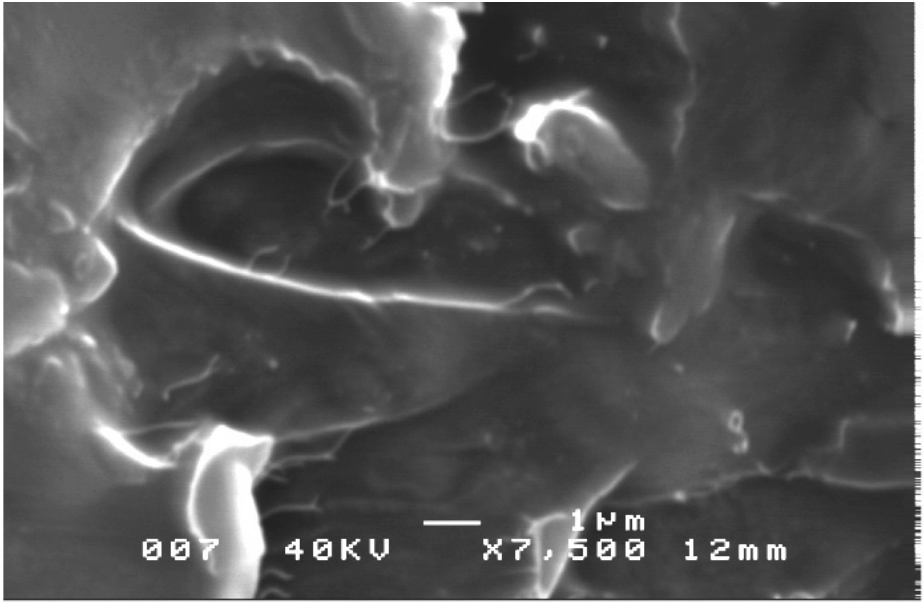

Figure 1, Biercuk et al.

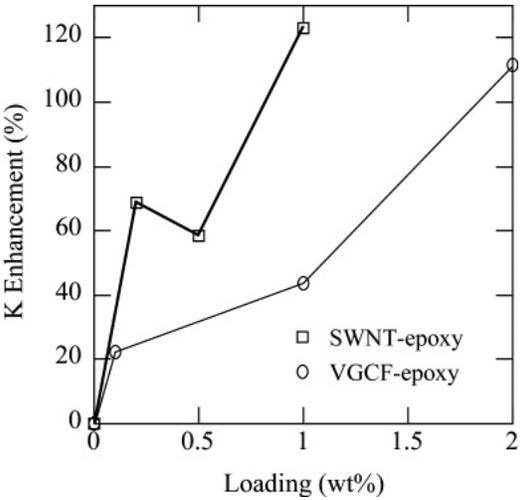

Figure 2, Biercuk et al.

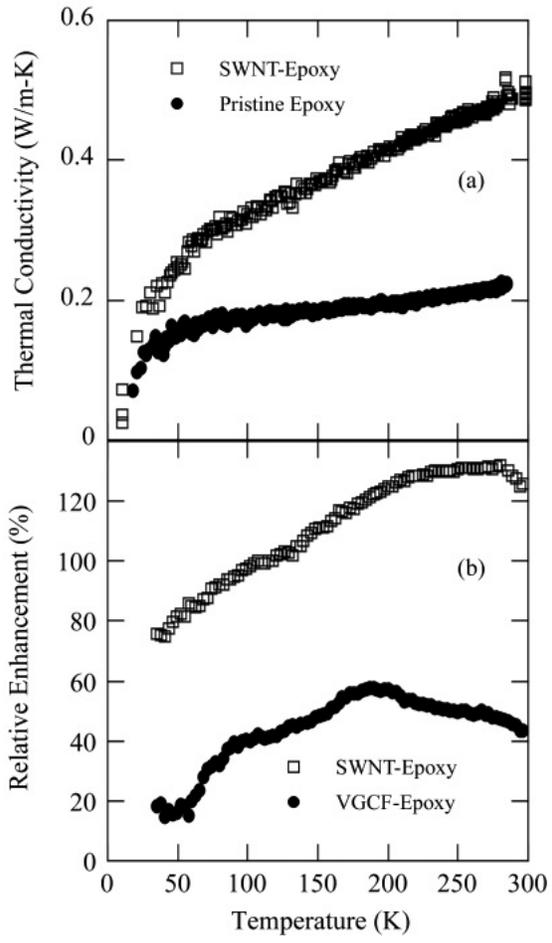

Figure 3, Biercuk et al.

[1] L.X. Benedict, S.G. Louie, M.L. Cohen, Solid State Commun. **100**, 177 (1996).

[2] J. Hone, B. Batlogg, Z. Benes, A.T. Johnson, J.E. Fischer, Science **289**, 1730 (2000).

[3] S. Berber, Y.K. Kwon, D. Tomanek, Phys. Rev. Lett. **84**, 4613 (2000).

[4] J. Hone, M. Whitney, C. Piskoti and A. Zettl, Phys. Rev. B Rapid Comm. **59**, R2514 (1999).

[5] J. Hone, M.C. Llaguno, N.M. Nemes, A.T. Johnson, J.E. Fischer, D.A. Walters, M.J. Casavant, J. Schmidt, R.E. Smalley, Appl. Phys. Lett. **77**, 666 (2000).

[6] P. Kim, L. Shi, A. Majumdar, and P. L. McEuen, Phys. Rev. Lett. **87**, 215502 (2001).

[7] R. Haggenmueller, H.H. Gommans, A.G. Rinzler, J.E. Fischer, and K.I. Winey, Chem. Phys. Lett. **330**, 219 (2000).

[8] L. Vaccarini, G. Desarmot, R. Almairac, S. Tahir, C. Goze, and P. Bernier, p. 521, Proceedings of the XIV International Winterschool, Kirchberg, Tyrol, 2000, edited by H. Kuzmany, J. Fink, M. Mehring, and S. Roth (AIP Conference Proceedings 544, Woodbury, New York).

[9] P. Ajayan, L. Schadler, C. Giannaris and A. Rubio, Advanced Materials **12**, 750 (2000).

[10] L.S. Schadler, S.C. Giannaris and P.M. Ajayan, Appl. Phys. Lett. **73**, 3842 (1998).

[11] H.Z. Geng, R. Rosen, B. Zheng, H. Shimoda, L. Fleming, J. Liu, and O. Zhou, unpublished.

[12] M. J. Bronikowski, P. A. Willis, D. T. Colbert, K. A. Smith, and R. E. Smalley, J. Vac. Sci. Technol. A, 19(4), 1800-1805 (2001).

[13] W. Zhou, Y.H. Ooi, R. Russo, P. Papanek, D.E. Luzzi, J.E. Fischer, M.J. Bronikowski, P.A. Willis, and R.E. Smalley, Chem. Phys. Lett. **350**, 6 (2001).

[14] Physical properties of Pyrograf-III were found on the Applied Sciences website: www.apsci.com.

[15] Solvent evaporation was considered complete when the remaining sample mass was within 0.1 % of the original mass of resin plus carbon material.

[16] M.C. Llaguno, J. Hone, A.T. Johnson, and J.E. Fischer, AIP Conference Proceedings **591**, 384 (2001).
5/13/02     Biercuk et al.     11